# Somatosensory Activation and Attentional States in Creative Making

Katherine Rees
University for the Creative Arts
katie.rees@uca.ac.uk

## ABSTRACT

The methods for capturing the creative process come with associated tensions around memory recall, articulation, and communication during the act of making, as well as how to record these considerations. This paper has a twofold purpose: first, to offer an example of a mixed methodology, drawn from dance anthropology, sensory ethnography, and design, that applies embodied methods as an alternative for documenting creative making. Specifically, this incorporates the researcher-as-participant and the collation of fieldnotes, embodied knowledge/movement recall, with notation forms, and participant interviews. These are existing methods in dance anthropology; however, using them alongside exploratory prototyping and workshop approaches broadened this work into transdisciplinary practice. Second, it discusses the activation of somatosensory systems through wearable technology and the facilitation of heightened sensory awareness for the practitioner, leading to a subsequent ability to focus on creative decisions linked to reflection and metacognition.

The term metacognition refers to "…reflection on and regulation of one's cognitive activities" [1]. Dance improvisation requires decision-making in the moment, and theories of embodied cognition [2] and 'thinking through the body' [3] can therefore be considered involved in this type of dance-making. However, improvisation in creative ideation can also generate movement ideas that are developed and evaluated, which is associated with metacognition as an overarching cognitive process. In this paper, I discuss how changing working relationships with creative digital tools are involved with varied attentional states and how these can be developed through heightened kinaesthesia through somatonsense activation. Attention, in this dynamically shaped technologically mediated situation, is less to do with an individual selecting to concentrate on one aspect of the stimuli, but instead a manner to deal with sensory information, processing sounds and sensation, and how to interpret and make meaning from this in creatively moving the body. By acknowledging different states, we can further consider how to capture and record reflection and metacognition in creative action.

## 1 Introduction

In investigating creative movement making, this paper draws on research involving dancing with and observing others using body-worn technology in The Digital Touch Project [4, 5]. This project employed digital touch stimuli generated by a bracelet that emits loops of varying vibrations via a haptic motor controller and a second piece of wearable technology: an inflatable silicone waistband powered by a motor pack applying a different sensation (pressure) on the core of the body. Workshops ranging from 4 to 9 hours were organised, focusing on the existing practices of the six professional dancers and choreographers. The participants were aged between 21 and 45 years, and had varied movement backgrounds, including ballet, contemporary, circus, breakdancing, classical Indian dance, martial arts and somatic techniques.

An inductive approach to practice-led research supported iterative prototype development and project-based qualitative data collection. My background in contemporary dance and the use of choreographers and dancers as participants in workshops enabled an exploration of my own role as the researcher and prototype developer. From facilitator to learner, to co-participant working in the space, I was able to develop an empathetic understanding of the real-time interaction.

To aid the discursive links between creative making and the creator's awareness of their creative decisions in an embodied manner, the following three sections will consider the involvement of attention and sensory information in the act of making, followed by explanations of embodied research in dance and scholarship on kinaesthesia and then a detailed description of the project methodology and findings.

## 2. Attention and awareness in creative making

Tim Ingold [6], in his anthropological discussion of skilled tool use, considers ongoing, low-level attention and the acknowledgement of affordances in creative making rather than automatic, unconscious movement. Lambros Malafouris's theory of Material Engagement [7] in cognitive archaeology considers the flow between the material, the maker, and the thinking process. In their journal article, Malafouris and Koukouti [8] define this as 'haptic attentive unity', further theorising the interaction between maker and material and the process of creative decision-making in the moment.

In the design process, the work of Wilde [9] and Nunez-Pacheco and Loke [10] considers the somatic register of the body and multisensorial experience in design ideation. The latter research proposes that the multimodality of experience is interconnected with participants' imagination and memory in the creative process. By employing embodied experience and somatic techniques that facilitate internal focus, designers investigate how individuals construct meaning through sensory experience. Wilde's considers how attentional qualities are directed toward the body through technological engagement. Through her practice-based research, she examines how interactive body-worn systems extend perception, imagination, and the bodily experience of the world through the body, and how these processes may alter relationships to the body and to body-worn technologies.

Within the literature on dance and technology, there is a growing body of work that considers the interaction between the dancer and the material, enabling multiple states of presence [11], and that multisensorial experiences of using technology in choreography necessitate a restructuring of senses. Davidson explains, "A mediated body emerges as a hybrid, multileveled response to interoceptive and exteroceptive stimuli." [12]. Subsequently, the need to process sensory information through interactive environments and digital tools enables the development of skills in attending to multiple sources of information simultaneously. This notion of different states of presence can be linked to attention.

Ingold proposes that finely tuned modes of attention can explain the apparently automatic or unconscious actions observed in creative tool use, whereby awareness is foregrounded and guided by sensory information, enabling adaptation and responsiveness to environmental changes [6]. Ingold does not argue that the tool becomes entirely backgrounded during assimilation; he suggests that practitioners employ a form of indirect attention to both material and environment throughout the process. This resonates with De Spain's concept of "decentred intentionality" [13] in his discussion of improvisation within interactive environments, operating via a process of filtering information/thoughts.

Katherine Hayles introduces the concept of "through/against/with/alongside" to theorise the extension of cognition through technological objects and the evolution of metacognitive capabilities via co-development with tools [14]. Both Hayles and Malafouris examine the symbiotic influences present in these interactions. Hayles emphasises the role of cognitive load in thinking through tools to account for the apparently unconscious experience, while Malafouris interprets the artist's responsiveness to both material and imagined affordances and influences as constituting consciousness itself [7].

Similar to Hayles, Gabora explains how multiple environmental information sources affect cognitive capacity, resulting in varied attentional states. Her rationale is that increased sensitivity to environmental features engages more memory locations, so a broad activation function may manifest as defocused attention or heightened sensitivity [15]. For instance, attentional change can be interpreted as a means of sensing and assimilating information, as described by Myers and Dumit's "creative mode of attention" [16]. Their use of time-lapse film loops and 3D immersive environments posits that the interaction between a participant and technology reconfigures understanding of data and phenomena.

Whether attention states and responsiveness are due to cognitive load or to a developed ability of working with tools (digital or otherwise), several scholars across disciplines describe a similar experience in the process of making. Before considering this specific research on somatosensory wearable technology, the context of bodily awareness in dance and its connection to embodied research will be outlined.

## 3. Embodied Research

Among those in dance ethnography who use analytical systems of movement analysis alongside their physical knowledge of learning dances [17], some scholars have discussed embodied knowledge in their research. Physical engagement has been considered to "… yield deeper levels of understanding and insight through engaged practice on different levels of immersion." [11] and by doing so, acknowledges the implications for the researcher as a participant. By bringing the "body as a lived reality" [18] to the forefront, we are confronted with the acceptance of our own training, habits, and socio-cultural influences. Andree Grau highlights how reflexivity can be a means to confront the researcher's biases and frames of reference [19]. Exploring sensory opportunities for understanding others' experiences should be considered alongside the verbal and non-verbal exchanges in the researcher-participant dialogue and the co-construction of meaning through this process [20]. Sarah Pink's [21] sensory ethnography further exemplifies the usefulness of participation in the research and how this approach can assist the researcher in establishing connections between experiences and the analytical processes.

## 4. Kinaesthetic awareness

Kinaesthetic awareness is a term in dance that describes the “double act of moving and feeling oneself move” [22]. As Sklar explains, the lack of widespread use of kinaesthesia or kinaesthetic awareness has been attributed to the historical preference for the five senses and due to the complexity of communicating this experience to others. Importantly for the focus of this paper, kinaesthesia in research both enables recognition in the moment and is revisited in data analysis through the embodied knowledge [22]. Bodily awareness in the process of action can be cultivated and developed through dance training and somatic practices.

Utilising kinaesthetic awareness in this research emerged from my own explorations with the prototypes and from learning movement phrases from the participants. The dual perspective provided by kinaesthetic awareness, in addition to bodily knowledge gained in demonstrating or copying another, provides an interpretative bridge between what we understand through our bodies and what we communicate to others. The key to metacognition lies in participation, sharing, and the ability to decentre attention and consider one's approach in the moment of creation.

## 5. Prototypes and Workshops

Exploratory prototyping and user testing yielded two prototypes. Through iterative development, the choice of passive touch to the body and different materials to vary the sensation was refined. A bracelet with a vibration motor and controller was developed with a coded loop of varied haptic sensations, worn on the wrists. The second prototype, combining a silicone waste band with a motorised pump, enabled inflation and deflation and provided sensation at the centre/core of the body. Utilising two prototypes alongside the choreographers' and dancers' existing practices in the project, presented an opportunity to study their use across different dance genres. It also allowed for extended time with the prototypes to be observed and analysed [4,5].

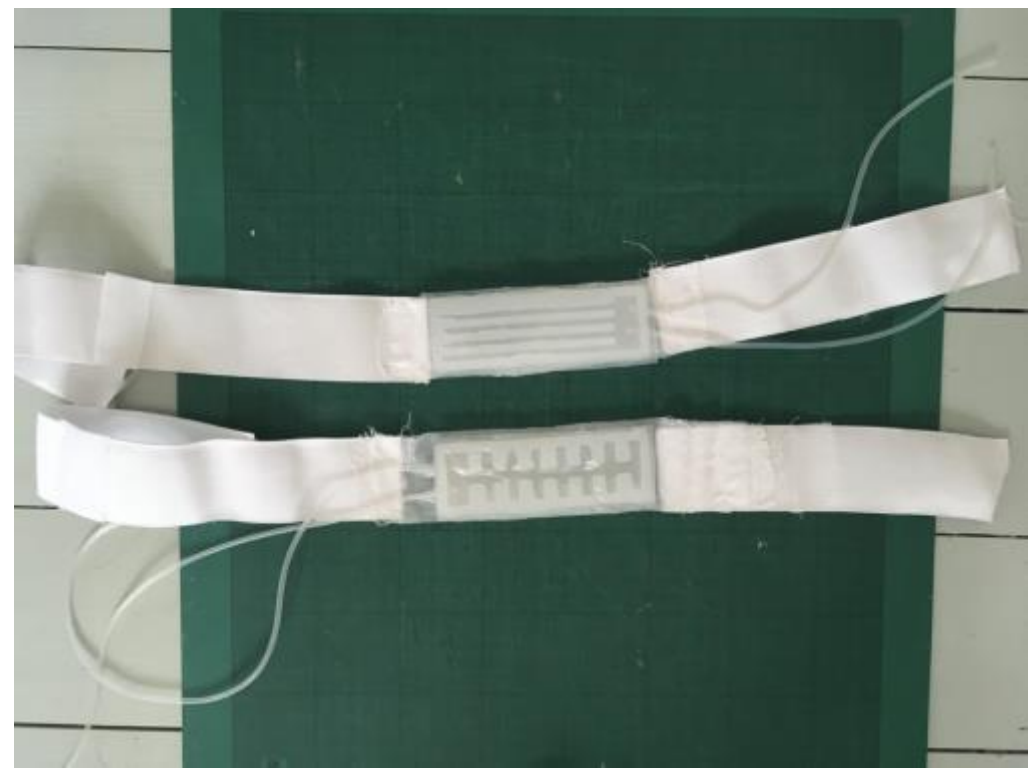

**Figure 1. Silicone inflatable belt for the pump prototype**

Two cameras were set up in the space to record the workshop, and this footage was used to interpret the movement generated. At the request of some participants, a playback of the video footage was used to discuss the movement. Recorded movement footage via video camera was notated using a modified version of Laban Movement Analysis, with categories of Body, Space, and Effort to describe the movement observed. Researcher fieldnotes and participant interviews were transcribed and coded, and this information was used to populate a matrix for each participant. An additional matrix collated the participants as a whole group and identified movement trends and qualitative themes via the analysis. This combination of mixed methods facilitates understanding of subjective experience by enabling participants to share and review their experiences during interviews, physically revisit aspects of improvisation, and articulate their intentions and responses to the stimuli.

## 6. Findings

Longer timeframes of prototype use with experienced dancers and choreographers provided evidence of different working relationships, as explained in Hayles's [14] theory of working with/through/against and alongside technology. Her discussion of technogenesis considers how media and technologies shape our cognition and behaviour in mutual ways. For example, the changes in reading skills and attention brought about by digital applications and tools. In this research, the theory has been applied at the micro-level to examine the opportunities and utilisation of digital touch in creative ideation and movement.

As detailed in Rees [4, 5], the repetitive sensation, both on the surface of the skin and resonating further into the body, from the prototypes disrupted a typical kinaesthetic awareness, extending this to further interoception (internal bodily sensations such as digestion, heart rate, etc.) and exteroception. A more directed attention at the beginning of working with the prototypes was witnessed; however, with further use, this evolved. In creative exploration, the desire to break free or to divert from the dominance of the prototype code was noticed in the interplay between syncing with and ignoring the rhythmic structure. Anticipation of the sensation/sound of digital touch manifested different working relationships with the prototypes, involving both passive and sometimes intentional active readjustments to the sensation and sound patterns. Attention states reflecting working through/against/with/alongside were linked to the participant's stance on the felt stimuli.

With extended use of the prototypes, the participants responded by devising rule-based tasks that disrupted the tendency to synchronise with sensation, thereby introducing an additional layer for exploration and experimentation. The manner in which participants established restrictions aligns with Western dance theatre practices for problem-based creation, as discussed by Cveji [23]. Her research examines the use of rules and problem-based frameworks to disrupt and fracture habitual patterns, fostering new approaches to creating and perceiving movement. This pathway encourages a more directed movement approach during the improvisation, working in opposition to aspects of sensation and sound or divergently to explore movement material. The process of engaging with the prototypes alternates between direct interaction, opposition, and collaborative or co-existent modes. The development of attentional states emerges from the active exploration and reciprocal shaping among prototypes, imagined and real environments, and the body.

Although the duration of prototype use and familiarisation was a factor, the prototypes did not facilitate straightforward mediation because they consistently produced a tactile sensation on the skin. The dancers' or choreographers' use of the prototypes is closely linked to attention and intention during movement generation.

> "It's got the feeling of listening to something. That was the sensation I had when doing it, that paying attention. …" (Participant 12 using bracelet vibration prototype in improvisation, Interview 1) [4].

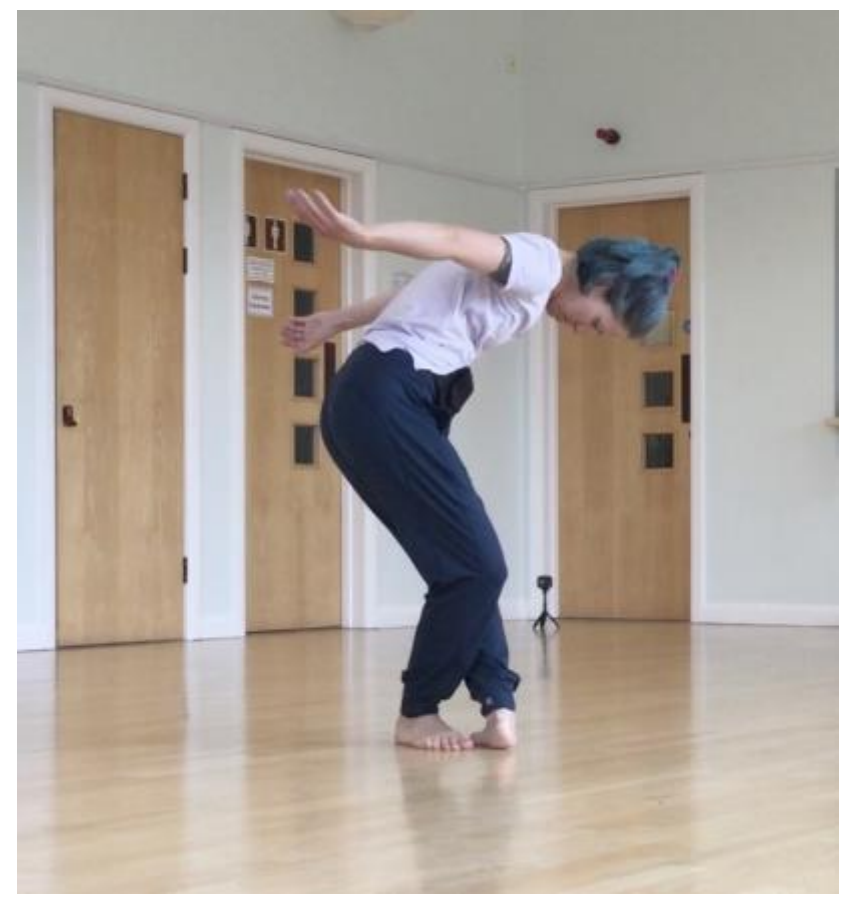

**Figure 2. Participant improvising with the pump prototype attached to the waist.**

Receiving information—conceptualised as 'hearing'—is central to the proposed modes of attention in working *through* the digital touch. In contrast, an *against* relationship involves active listening, but decisions are guided by predetermined rules or agendas. When working *with* the prototype, the relationship incorporates both listening and hearing, leading to selective synchronisation. In the final state, the participant and prototype operate *alongside* each other, combining hearing and listening to facilitate noticing and a more balanced body-technology relationship that supports metacognition.

In some cases, digital touch stimuli disrupted this process during participants' movement creation. For instance, during improvisation, several participants in both projects experienced a divergence between their intentions and the resulting movement. Movement analysis and interview data indicated a 'stop-start' quality, attributed to the anticipation of different vibration patterns and the ongoing feedback loop. The regulation of pace and the dynamic qualities of slowing and flattening the movement were witnessed during the use of the binary (inflate-deflate) of the pump prototype. Although negative feedback or readjustment occurred between the code's aesthetics and the moving body, this process can be interpreted as mutual adaptation and selective synchronisation. Data indicate that, with extended use of the prototypes, most participants progressed beyond simple reactions to stimuli and developed new relationships characterised by varied modes of attention. When working *against* the prototypes, observed movements demonstrated resistance to rhythm, dominance of the prototype's location, or reliance on rule-based strategies to increase improvisational complexity. In *against* or *alongside* relationships, the prototypes exerted less influence on movement, particularly when structuring or editing phrases, allowing for greater diversity in movement action, dynamics, effort, and levels. Participants 14 and 15, whose existing choreographic practice incorporated meditation and mindfulness, unsurprisingly adopted attentional strategies running *alongside* the digital touch stimuli. However, Participants 12 and 13 engaged in a process similar to metacognition by noticing and filtering their thoughts and working in a decentred attention with the technology.

> "I found myself definitely aware and kind of moving with or moving around the rhythms. Certainly not directly in relation to them, but they were very much supporting the temporality of my movement and my experience. And also, a feeling of it being sometimes a fulcrum for the way's things were organising. So, it would every now and again become the centre of gravity for my perception and awareness. (Participant 14, Interview 2)"[4]

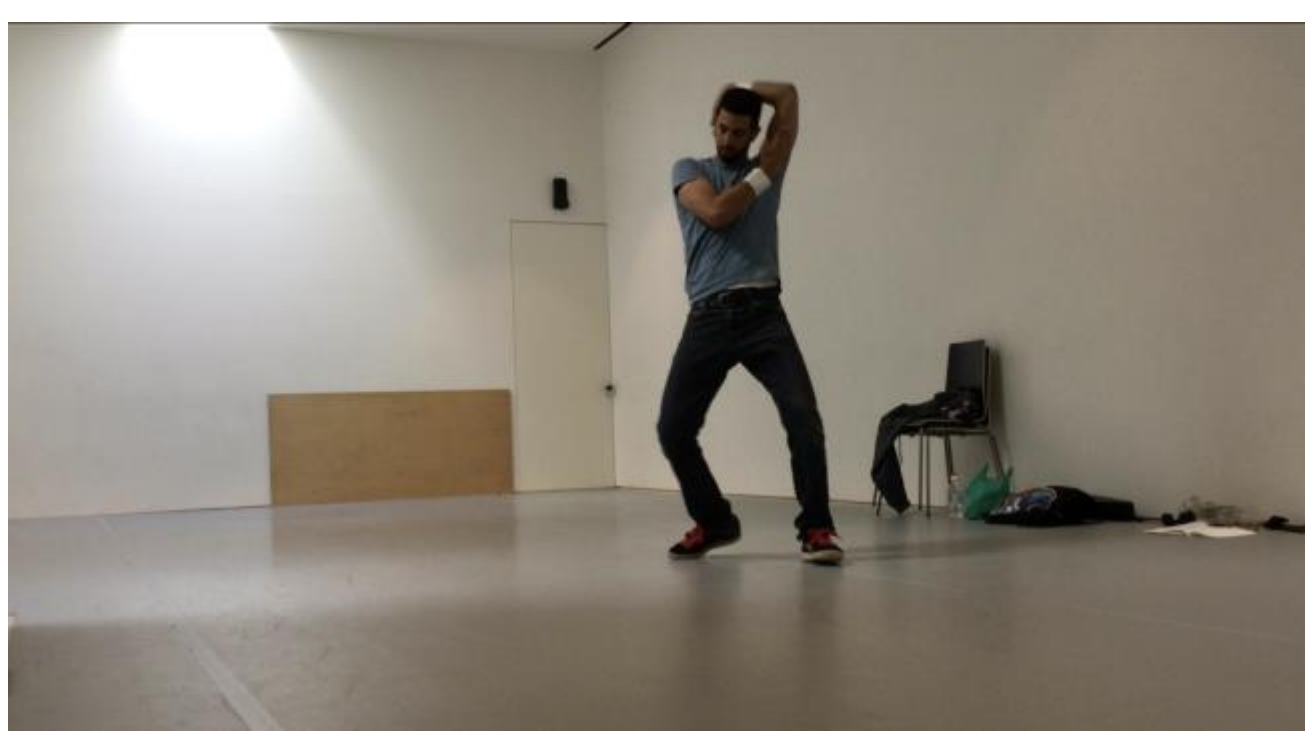

**Figure 3. Participant improvising with two vibration prototypes worn at the wrists.**

More than three decades ago, Varela, Thompson, and Rosch's 'Embodied Mind' [24] advocated for incorporating mindfulness, rooted in Buddhism, through diverse attentional states in cognition. In the revised edition, Thompson clarifies the relationship between mindfulness and enaction, explaining that meditation is a practised skill that allows behaviours/states to engage with thinking through the act of doing [25]. Despite two of the participants ability to utilise skills of meditation in working with the prototypes, the qualitative data from both projects identified codes such as 'foci', 'fulcrum', 'amplification', and 'heightened awareness', which support the notion that working with digital touch stimuli can facilitate a pronounced and simultaneous 'doing' and 'noticing' across the participant group.

> "I just feel it doesn't necessarily make me move in a different way, it somehow shines light on a brighter area of my understanding of what I am doing in space with X. It's like if I was in a room and they only followed me with a torch but wearing… like… six torches. (Participant 15, Interview 2)" [4]

The ability to notice, filter, and change creative strategies can be viewed as beyond that associated with reflection-in-action, as it encompasses an equitable acknowledgement of stimuli by working 'alongside'. This is not suggestive of 'on the spot' decisions to solve problems, but an ability to self-monitor and guide embodied cognition.

## 7. Conclusion

Findings indicate that the use of the prototypes promoted a heightened appreciation of the body. The body worn technology offered contact through passive touch sensation and subsequently served as a reference point. Notably, somatoperception enabled participants to experience their bodies differently, with the coded loop offering repeated sensations. This process bridged the participants' kinesphere, fostering a new awareness of space and time that included both interoception and exteroception.

Analysis of the workshops revealed changes in how participants engaged with the prototypes, as discussed through the framework of working 'through,' 'against,' 'with,' and 'alongside' [4]. Within this context, the aesthetics of the loop were interpreted as direct responses to the code, characterising the relationship as working *through* the prototypes. Participants described their responses as pure reactions to the stimuli at the beginning of their use of the prototypes. Some moved towards working *with* the prototypes, a process supported by the theory of synchronicity, which suggests active listening and passive hearing in comprehending digital touch. A more parallel approach, described as working *alongside*, utilises an attention state linked to metacognition. In creative exploration, some participants intentionally diverged from established patterns or created new rules to disrupt their responses, a process working *against* the suggested rhythms and sensations of the prototype code [4,5]

The combination of movement and tactile sensation suggests a new understanding of the body. This comprehension is facilitated by developed kinaesthesia and heightened awareness of the fluid boundaries between the body's interior and exterior. Recommendations for using other technologies that stimulate somatosensory perception would be worth exploring to further examine their potential for attentional states and for identifying the optimal conditions for metacognition in creative action. In addition, opportunities for embodied research, in which the process is explained through watching, sharing, and learning, would be recommended to support the connection between what is felt, considered and materialises.

Future research in this area would be enhanced by advances in neurotechnology and the development of mobile electroencephalography (EEG), which could yield more accurate brain data in action. This would create additional means to record different states and body awareness by collecting and communicating real-time data on the nervous system and cognition.

## ACKNOWLEDGMENTS

This work was supported by The British Academy – Talent Development Award 2024-2025 and Studio Wayne McGregor and London College of Fashion PhD Fund.

## REFERENCES

[1] Izabela Lebuda and Mathias Benedek. 2023. A systematic

framework of creative metacognition. Physics of Life Reviews, 46:161-181. DOI:10.1016/j.plrev.2023.07.002.

[2] Shaun Gallagher. 2017. Enactivist Interventions: Rethinking the Mind. Oxford. Oxford University Press.

[3] Alva Noë. 2004. Action in Perception. Cambridge. Massachusetts. The MIT Press.

[4] Katherine Rees. 2022. Somatosensory wearable technology in the dance making process. Ph.D. Thesis. London College of Fashion. University of Arts London. London.

[5] Katherine Rees. 2023. Using digital touch in the process of improvising and the implications for kinaesthesia. Journal for Dance and Somatic Practices. 15(2). 213 – 229. DOI: 10.1386/jdsp_00110_1

[6] Tim Ingold. 2011. Being Alive: Essays on Movement, Knowledge and Description. Routledge. Abingdon

[7] Lambros Malafouris. 2013. How Things Shape the Mind. MIT Press

[8] Lambros Malafouris and Maria-Danae Koukouti. 2022. Where the touching is touched: The role of haptic attentive unity in the dialogue between maker and material. Multimodality & Society, 2(3). 265–287. DOI: 10.1177/26349795221109231

[9] Danielle Wilde. 2011. Swing That Thing: moving to move. The poetics of embodied engagement, Ph.D. thesis. Monash University, Melbourne, Australia

[10] Claudia Nunez- Pacheco and Liane Loke. 2017. Tacit Narratives: Surfacing Aesthetic Meaning by Using Wearable Props and Focusing. Proceedings TEI 2017. Yokohama. Japan. March 20–23. DOI: 10.1145/3024969.3024979

[11] Andrea Davidson. 2013. Somatics: An orchid in the land of technology. Journal of Dance & Somatic Practices, 5(1). 3 –15.

[12] Andrea Davidson. 2016. Ontological shifts: Multi-sensoriality and embodiment in the third wave of digital interfaces, Journal of Dance & Somatic Practices. 8(1). 21–42.

[13] Kent De Spain. 2011. Improvisation and intimate technologies. Choreographic Practices, 2.25–42.

[14] Katherine Hayles. 2012. How We Think: Digital Media and Contemporary Technogenesis. The University Press of Chicago. Chicago.

[15] Liane Gabora. 2000. Toward a Theory of Creative Inklings. In Art, Technology, Consciousness: Mind@large, Roy. Ascott (ed.) Bristol: Intellect.159-164.

[16] Natasha Myers and Joe Dumit. 2011. Haptic Creativity and the Mid-embodiments of Experimental Life. In A Companion to the Anthropology of the Body and Embodiment. Frances Mascia-Lees (ed.) Oxford: Blackwell Publishing. 239-261.

[17] Adrienne Kaeppler.1972. Method and Theory in Analyzing Dance Structure with an Analysis of Tongan Dance. Ethnomusicology. XVI (2). 173-217.

[18] Helen Thomas and Jamilah Ahmed. Eds. 2004. Cultural bodies: Ethnography and theory. Malden, MA: Blackwell Publishing. 285

[19] Andre Grau. 2007. Dance Anthropology and Research through practice. Retrieved from: www.academia.edu/199889/Dance_anthropology_and_research_through_practice (Accessed 12 May 2026).

[20] Susanne Ravn and Helle Ploug Hansen. 2013. How to explore dancers' sense experiences? A study of how multi-sited fieldwork and phenomenology can be combined'. Qualitative Research in Sport, Exercise and Health. 5(2). 196-213. DOI: 10.1080/2159676X.2012.712991

[21] Sarah Pink. 2013. Doing Sensory Ethnography. London: Sage.

[22] Deidre Sklar. 2000. Reprise: On Dance Ethnography. Dance Research Journal. 32(1). 70–77.

[23] Bojana Cveji. 2017. Problem as a Choreographic and Philosophical Kind of Thought. In The Oxford Handbook of Dance and Politics. Rebekah Kowal, Randy Martin, and Gerald Siegmund (eds.). New York: Oxford University Press. Chapter 11.

[24] Francisco Varela, Evan Thompson. and Eleanor Rosch .1992. The Embodied Mind: Cognitive Science and Human Experience. Cambridge, MA: The MIT Press.

[25] Francisco Varela, Evan Thompson. and Eleanor Rosch. 2017. The Embodied Mind: Cognitive Science and Human Experience. Cambridge, MA: The MIT Press.